# Comment on "Casimir torque on two rotating plates"

A.A. Kyasov and G.V. Dedkov[1]

Nanoscale Physics Group, Kabardino-Balkarian State University, Nalchik,360004, Russian Federation

We discuss the recent result by Xiang Chen (Int. J. Mod. Phys., 2013) and show that the formula obtained for the Casimir torque in the systems of two isotropic rotating particles (plates) is incorrect.

It is known that Casimir torque on two anisotropic or birefringent plates is caused by vacuum fluctuations and boundary conditions [1]. Recently, the Casimir torque in the systems of two isotropic rotating particles and two rotating plates with one rotation axis but with different temperatures and/or different angular velocities with respect to vacuum has been calculated [2]. The particles are characterized by polarizabilities, while the plates are considered within a simplified model, where the Casimir torque is calculated through the torque between the two elementary rotating dipole particles. The author has come to several surprising conclusions, particularly that even for two nonrotating particles but with different temperature a torque can also be generated. Based on the formula for the torque between two rotating particles, he claims that similar to the two-particle system, there exists a temperature "induced" torque due to the difference in temperature on each plate even without rotation. In this short communication we show that these conclusions are physically incorrect.

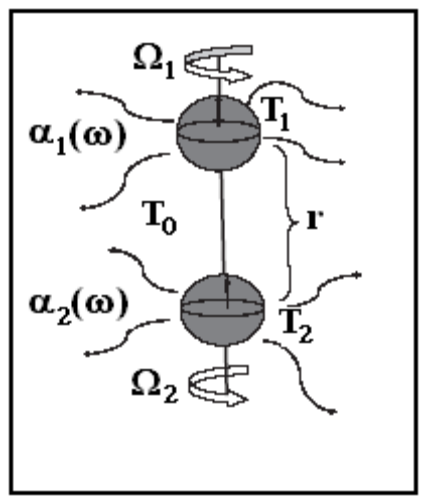

**Figure.** The system of two neutral rotating particles with polarizabilities $\alpha_1(\omega), \alpha_2(\omega)$, temperatures $T_1, T_2$ and angular velocities $\Omega_1, \Omega_2$ embedded in the vacuum with temperature $T_0$.

---

[1] Corresponding author e-mail: gv_dedkov@mail.ru

The figure shows the system being considered, relating to the configuration of two fluctuating dipole particles embedded in vacuum. According to the main result of [2] and assuming that the particles are isotropic, the torque on particle 1 is given by (Eq. (37) in [2]), namely

$$M_1 = \frac{4\hbar}{3\pi c^3} \int_{-\infty}^{+\infty} d\omega\, \omega^3 \operatorname{Im}\alpha_1(\omega_+)[n_1(\omega_+) - n_0(\omega)] +$$
$$+ \frac{2\hbar}{\pi r^6} \int_0^\infty d\omega \operatorname{Im}\{\alpha_1(\omega_+)\}\operatorname{Im}\{\alpha_2(\tilde{\omega}_+)\}[n_1(\omega_+) - n_2(\tilde{\omega}_+)] \quad (1)$$

where $\omega_+ = \omega + \Omega_1, \tilde{\omega}_+ = \omega + \Omega_2$, and

$$n_i(\omega) = \frac{1}{\exp\left(\dfrac{\hbar\omega}{kT_i}\right) - 1}, \quad i = 0,1,2$$

The first term in (1) relates to the interaction of particle 1 and the vacuum, and the second term relates to the interaction of particle 1 and particle 2, since it depends on the excitations $n_1(\omega_+)$ and $n_2(\tilde{\omega}_+)$. The first term in (1) coincides with the expression obtained in [3], having a purely relativistic character, and it is omitted in our further analysis. The second term in (1) corresponds to only nonrelativistic approximation. We can see that the net torque corresponding to this second term would not vanish even if both particles are at rest ($\Omega_1 = \Omega_2 = 0$). It is this result has led the author to the similar conclusion when calculating the torque on two nonrotating plates.

Prior to [2], the problem of torque on two rotating particles has been considered in our works [4--6], but we used the problem statement (both relativistic and nonrelativistic) when the second particle is at rest with respect to vacuum ($\Omega_2 = 0$). According to our nonrelativistic result [4,6], the torque on particle 1 has the form ($\omega_\pm = \omega \pm \Omega_1$)

$$M_1 = \frac{2\hbar}{\pi r^6} \int_0^\infty d\omega \operatorname{Im}\alpha_2(\omega) \cdot \{\operatorname{Im}\alpha_1(\omega_+)[n_1(\omega_+) - n_2(\omega)] - \operatorname{Im}\alpha_1(\omega_-)[n_1(\omega_-) - n_2(\omega)]\} \quad (2)$$

As we can see from (2), in the case of two nonrotating particles ($\Omega_1 = \Omega_2 = 0$), $M_1 \equiv 0$ irrespectively of the temperature and polarizability of each particle, in contrast to (1), from which we obtain (see Eq. (38) in [2])

$$M_1 = \frac{2\hbar}{\pi r^6} \int_0^\infty d\omega \, \mathrm{Im}\{\alpha_1(\omega)\} \mathrm{Im}\{\alpha_2(\omega)\}[n_1(\omega) - n_2(\omega)] \tag{3}$$

In our opinion, this result is incorrect and physically senseless, since $M_1 \neq 0$ at $T_1 \neq T_2$.

From this, contrary to the further conclusion of the author [2], the presence of heat exchange between the two nonrotating particles embedded in vacuum (background radiation) can not violate spatial isotropy of vacuum.

Correct expression for the torque on particle 1 when both particle 1 and particle 2 rotate, can be obtained in line with our calculation in [4]. The result is given by

$$M_1 = \frac{2\hbar}{\pi r^6} \int_0^\infty d\omega \cdot \begin{Bmatrix} \mathrm{Im}\,\alpha_1(\omega_1^+) \mathrm{Im}\,\alpha_2(\omega_2^+)[n_1(\omega_1^+) - n_2(\omega_2^+)] - \\ -\mathrm{Im}\,\alpha_1(\omega_1^-) \mathrm{Im}\,\alpha_2(\omega_2^-)[n_1(\omega_1^-) - n_2(\omega_2^-)] \end{Bmatrix} \tag{4}$$

where $\omega_1^{\pm} = \omega \pm \Omega_1$, $\omega_2^{\pm} = \omega \pm \Omega_2$.

From (4) it follows that in the case of two identical particles, the torque on particle 1 proves to be zero in two cases: i) at $\Omega_1 = \Omega_2 = 0$ and arbitrary temperatures $T_1$ and $T_2$; ii) at $\Omega_1 = \Omega_2$ and $T_1 = T_2$. If $\Omega_1 = \Omega_2$ but $T_1 \neq T_2$, the torque differs from zero. This fact is rather unexpected and surprising.

Relativistic calculation in the case where both particles rotate relative to vacuum, needs a special analysis. As to the two rotating plates, the problem is still more complicated and it is also a challenge for further investigation.